\documentstyle[12pt,aasms4]{article}

\slugcomment{Submitted to The Astrophysical Journal }

\lefthead{Marani et al.}
\righthead{Correlations Between Clusters and GRBs}

\begin{document}

\title{On Suggestive Correlations Between GRBs and Clusters of Galaxies}

\author{Gabriela F. Marani, Robert J. Nemiroff\altaffilmark{1,2},}
\affil{Center for Earth and Space Research, George Mason University,
Fairfax, VA 22030}

\author{Jay P. Norris and Jerry T. Bonnell\altaffilmark{3}}
\affil{NASA Goddard Space Flight Center, Code 660.1, Greenbelt, MD 20771}

\altaffiltext{1}{Visiting Scientist, Compton Science Support Center, NASA GSFC}
\altaffiltext{2}{Michigan Technological University, Houghton, MI 49931}
\altaffiltext{3}{Universities Space Research Association}

\begin{abstract}
Recent claims of angular correlations between gamma-ray bursts (GRBs) and
clusters of galaxies are evaluated in light of existing but previously
uncorrelated GRB positional data. Additional GRB data sets we use include
sub-samples of soft BATSE 3B bursts, bursts located by the Interplanetary
Network (IPN), and GRBs localized by COMPTEL. We confirm a previously
reported excess by Rood and Struble (1996) of the 185 rich, nearby clusters
of galaxies (Abell, Corwin, and Olowin 1989, ACO) in the 1-$\sigma$ error
circles of 74 BATSE 3B positions, but find a typical correlation strength
of only $~$2.5-$\sigma$ for typical sub-samples. However, none of the 185 ACO
clusters occur in the 1-$\sigma$ error boxes of 40 IPN GRBs or 18 COMPTEL
GRBs.  When all ACO clusters are correlated with BATSE 3B GRBs however, we
find an increasingly strong correlation for GRBs with decreasingly small
error boxes, reaching above the 3.5-$\sigma$ level. We also find a slight
excess of {\it soft} BATSE GRBs near the positions of 185 rich, nearby ACO
clusters, but the significance of the correlation averages only
$~$2.5-$\sigma$ for sub-samples delineated by softness. We caution that the
statistical significance of all these correlations is marginal, and so
conclude that the excess is at best only suggestive of a physical
association. Statistical fluke is still a strong possibility. BATSE could
confirm or refute such correlations in a 10-year lifetime.
\end{abstract}


\section{Introduction}
The nature and distance scale to gamma-ray bursts (GRBs) is still under
debate almost three decades after their discovery (Klebesadel, Strong \&
Olson 1973; Paczynski 1995; Lamb 1995). A significant correlation between
the angular positions of GRBs and clusters of galaxies would indicate a
setting for the burster objects.

A preliminary study with first year data from the Burst and Source
Transient Experiment (BATSE; ``1B data") on board the Compton Gamma Ray
Observatory (CGRO) by Howard et al. (1993) indicated no significant
correlation between GRBs and clusters from the catalog of Abell, Corwin, \&
Olowin (1989, ACO).  A lack of correlation was again found between BATSE
second year data (2B) and ACO clusters by Nemiroff, Marani, and Cebral
(1994).

A correlation at slightly over the 2-$\sigma$ level between ACO clusters
and GRBs at 10 degrees was reported by Cohen, Kolatt \& Piran (1994) in
BATSE 2B data.   A subsequent analysis by Rood and Struble (1996), however,
reported a 3-$\sigma$ correlation between BATSE 3B GRBs for which the
positional uncertainties are less than 1.685 degrees, and ACO clusters with
$R \ge 1, D \le 4$, where $R$ stands for richness class and $D$ stands for
distance class.

We examine here the correlative power of a wider data base of potentially
local GRBs with ACO clusters. Specifically, we cross-correlate GRB
sub-samples that give indications of following a Euclidean brightness
distribution (``Log N - Log P") and hence, in the cosmological distance
scale paradigm, would indicate that GRBs exist at a distance scale similar
to the ACO clusters (Wickramasinghe et al. 1993).  Sub-samples analyzed
include 3B bursts of soft spectra, 40 bursts from the IPN catalog, and 18
bursts with roughly degree-sized error regions detected by COMPTEL. We also
examine previous claims of GRB - cluster correlations, and re-evaluate the
significance of these claims.

In Section 2 we discuss the GRB samples used in more depth. In Section 3 we
discuss the computational method used to cross-correlate GRBs and clusters
of galaxies. In Section 4 we describe results, and in Section 5 we present
discussion and conclusions.

\section{Data}

GRB positional data were taken from three sources: the BATSE 3B catalog
(Fishman et al. 1994; Meegan et al. 1994), the IPN catalog of GRBs (Atteia
et al. 1987), and from GRB localizations reported by Kippen (1995) for
COMPTEL observations. BATSE 3B positional data has two sources of error:
random and systematic. A random error is associated with each GRB, whereas
the brightness-independent systematic error is listed as 1.6 degrees in the
3B catalog preamble. We compute the total error as the quadrature sum of
the random and the systematic.

The most accurate method of locating GRBs to-date is by using arrival time
information between three or more satellites located in different parts of
the solar system.  In the 1970s and 1980s, as many as nine such satellites
simultaneously recorded bright GRBs.  Collectively the GRB detectors on
these satellites are known as the Inter-Planetary Network (IPN; Cline 1992,
Hurley 1994). A catalog of 84 GRB positions was published by Atteia et al.
(1987).  Typical systematic errors for these ``IPN" GRBs are about 0.5
degrees; however, many times the error boxes were quite elongated.

The COMPTEL instrument has recorded more accurate GRB positions than BATSE
- typically locating GRBs in its field of view to about one degree
accuracy.  Few GRBs recorded by BATSE, however, are both bright enough and
positioned well enough for COMPTEL to record so accurate a position.
Kippen (1995), in his Ph. D. thesis, reports 18 GRBs which have total
positional errors of about 1.25 degrees or less.

Each GRB we use in the present analysis can be characterized by a
position and a 1-$\sigma$ error box in total positional uncertainty.
Although the distribution of positional errors does not usually follow a
Gaussian distribution, the 1-$\sigma$ error boxes for BATSE GRBs are
calibrated to include 68.3 \% of all GRB positions (Pendleton et al. 1995),
and we assume the same is true for other GRB catalogs as well.

Data on clusters of galaxies were taken from tables in ACO. An electronic
version of these tables was downloaded from the archives of the High Energy
Astrophysics Science Archive Research Center (HEASARC) in February 1996. We
considered 5250 ACO clusters in all, storing position, richness class,
distance class, and approximate redshift.  The primary sub-catalog of ACO
clusters described by Rood and Struble (1996) comprises 185 clusters with
richness class $R \ge 1$, and distance class $D \le 4$.  We assume the
published position of an ACO cluster to be exact.

\section{Correlation Methods}

Our goal is to estimate the probability of correlation between GRBs and ACO
clusters. We compute a correlation statistic as follows: Count the number
of ACO clusters that fall within the 1-$\sigma$ error circle of all of the
GRBs.  Demand that each GRB match {\it at most} one cluster (Rood, private
communication).  Throw 1000 sets of GRBs to random locations on the sky,
each set containing the same number of GRBs as in the actual GRB sample.
The error circles for these GRBs are chosen randomly from the actual GRB
error-box set. This ``bootstrap" method (Barrow et al. 1984; Simpson \&
Mayer-Hasselwander 1986), is particularly good at correcting for any ACO
clustering, since the same clustering would appear in the ``random"
samples. The randomly thrown GRBs are then ranked in terms of how many ACO
clusters were found within the error boxes.  We subsequently rank the
position of the {\it actual} number of ACO coincident clusters among the
``randomly correlated" ACO cluster list, and thus determine how distant
from the median it lies. We define a 1-$\sigma$ (2-$\sigma$, etc.) distance
from the median for ACO/GRB pair counts falling just outside a band
containing 68.3 \% (95.4 \%, etc.) pairs, centered on the median.

To search for a correlation, we first sorted the GRB list by some attribute
(softness, area of positional error box, etc.).  We then took the first GRB
on the list and computed the significance of correlation with rich nearby
ACO clusters by the above method. Next, we took the top {\it two} GRBs from
the list and again correlated them with ACO clusters.  We continued this
for the top three, four, and so on until the whole list was
cross-correlated with the ACO catalog.  This method generates an array of
correlation statistics and avoids reporting only a best correlation result.
As a control test, we also computed correlations for positions antipodal to
that of each GRB.

Each of these calculations is compared only with the null hypothesis: that
GRBs are {\it not} correlated with ACO clusters in any way.  It would be
interesting to compare the results with a model of GRB/ACO correlation.
However, no such definitive 3-D model exists, and any ``simple" testable
correlation would have to hypothesize a luminosity function for GRBs, to
hypothesize a resulting 3-D correlation function between ACO clusters and
GRBs, and, from these, find the detection efficiency for GRBs from more
distant clusters. We feel this is work for a more advanced project. We feel
that a simple correlation should first be established before such a project
is embarked upon.

\section{Results}

For the BATSE sample of 74 GRBs having error radii less than
1.685 degrees, defined and used in Rood and Struble (1996), we are able to
confirm the slight statistical excess of GRBs near ACO clusters reported by
Rood and Struble (1996).   Specifically, we found 7 GRBs within 1.685
degrees of 185 ACO clusters with $R \ge 1, D \le 4$. These GRBs are: (A496,
B2889), (A1749, B1711), (A3379, B2700), (A754 , B2603), (A3651, B1709),
(A3381, B3138), and (A566, B2855), where A stands for ACO cluster catalog
number and B stands for BATSE 3B catalog trigger number.  The expected
number of pairs predicted by our Monte Carlo runs was about 2.4, and we
estimate the significance of the result as between 2.7 and 2.9-$\sigma$.

To further test the Rood and Struble (1996) result, we computed the
correlation significance for the 185 bright, nearby ACO clusters with
incremental sub-samples of GRBs having increasingly larger error boxes.  We
found that the 74 best located GRBs used in the Rood and Struble (1996)
result sit near a maximum in significance, and about 2.5-$\sigma$ is a more
representative significance level for GRBs in this error box range.  The
highest peak was found for the 71 best located GRBs which yielded a
significance level of between 2.9 and 3.5-$\sigma$.

We selected from the IPN sample 40 GRBs with quoted 1-$\sigma$ error boxes
having areas less than or equal to the largest error area of BATSE 3B GRBs
used by Rood and Struble (1996).  We additionally required that no IPN
error box axis was greater than or equal to 1.685 degrees, the error box
radius used in Rood and Struble.  For these IPN GRBs, we
found zero ACO clusters with $R \ge 1, D \le 4$ inside these cumulative
error boxes.  Our Monte Carlo simulation of 1000 runs of 40 random GRB
positions found the following percentage results: (number of ACO
cluster /  IPN GRB pairs,
fraction of trials having exactly this number): (0, 0.767), (1, 0.200), (2,
0.030), (3, 0.004).  We thus find that the lack of bright, nearby ACO
clusters near IPN GRB positions is consistent with random GRB locations and
hence with no GRB - ACO cluster association.

The COMPTEL sample of 18 GRBs has quoted 1-$\sigma$ errors less than or
equal to 2.35 degrees.  We again found no ACO clusters with $R \ge 1, D \le
4$ inside these cumulative error boxes.  We again ran 1000 Monte Carlo
simulations using random GRB positions and found the following percentages
of GRBs: (number of ACO cluster / COMPTEL GRB pairs, fraction of trials
having exactly this number): (0, 0.721), (1, 0.229), (2, 0.045), (3,
0.006). We thus again find that the lack of bright, nearby ACO clusters
near COMPTEL GRB positions is consistent with random GRB locations and
therefore consistent with no GRB - ACO cluster association.

To check whether these slight excesses were anomalous, we computed
correlation strengths for {\it all} ACO clusters using incremental samples
of GRBs with increasingly larger error boxes.  The running correlation
strength is plotted against random location error in Figure 1. Note that in
addition to the random locations error, a systematic location error of 1.6
degrees is also reported. The vertical dashed line in Figure 1 marks the
number of GRBs for which Rood and Struble (1996) reported their main
correlation. Figure 1 shows a maximum correlation strength at the 27 best
located GRBs, where the correlation strength was between 3.562 and 4-
$\sigma$. For this set, 21 GRBs had an ACO cluster located inside their
1-$\sigma$ error box, while only 11.87 would have been expected from a
random distribution. From inspection of the Figure 1 we find that the
correlation strength drops off as bursts location accuracy decreases.

We note that each point in Figure 1 is {\it not} statistically independent
from others - each data point also represents all points to its
left.  We also note that the rising part of the Figure 1 curve -- for the
best located GRBs -- may be considered evidence against a statistical
association. In this region, random error is significantly less than
systematic error (1.6 degrees), so that a constant correlation fraction
might be expected.  This constant fraction might yields a rising
correlation strength as more GRBs are included.  But, more probably, the
analysis is just dominated by the inaccuracy of small number statistics in
this region.  Whether the shape of Figure 1 would be expected from a
GRB/ACO correlation depends exactly on the nature of the correlation.

We also tested soft BATSE GRBs for positional correlation with ACO
clusters. The softness parameter we used is defined as the sum of the
channel 3 and 4 fluences divided by the sum of the channel 1 and 2
fluences, as listed in the BATSE 3B catalog (Fishman et al. 1994; Meegan et
al. 1994), where the discriminator channel boundaries are 25, 55, 110, and
320 keV. Correlation results are shown in Figure 2. The two jagged lines in
Figure 2 show the maximum and minimum significance levels attributable to
the number of recorded coincidence pairs. From inspection of the figure we
see a modest excess of $R \ge 1, D \le 4$ ACO clusters near the
positions of soft BATSE GRBs. Although the correlation strength reaches as
high as 3-$\sigma$, inspection of the figure shows that 2.5-$\sigma$ is a
more fair representation of the correlation strength for soft GRBs.  When
all GRBs were considered, so that no softness selection existed in the
sample, the correlation strength dropped to about 1.6-$\sigma$.

In general, no significant correlations were found when antipodal positions
of GRBs were used as a control test in all of the above cases.  An odd
exception to this was for the 185 bright nearby ACO clusters when
correlated with the antipodal positions of the 900 to about 1100 best
located GRBs, where the correlation strength averaged about 2.5-$\sigma$
but peaked at over 3.5-$\sigma$.  The total correlation strength between
these ACO clusters and the antipodal positions of all 1122 BATSE 3B GRBs
was about 2-$\sigma$.  We can find no good reason to expect a correlation
with antipodal positions other than statistical fluke - we only report it
here to give the reader a feeling for how exhaustive an analysis was done
in generating the quoted results.

We also tested to see if correlations were a function of other variables.
We found a slight tendency for GRBs to correlate with rich ACO clusters
over poorer ones, and nearby clusters over further ones, but not at a level
of convincing statistical significance.

\section{Discussion and Conclusions}

The evidence for a correlation between GRBs and clusters of galaxies is
currently only marginal.  We add here, however, two new marginal results.
The first involves the best located GRBs when correlated with {\it all} ACO
clusters, and the second that involves soft GRBs.

We were able to confirm the excess of GRBs near rich, nearby ACO positions
first reported by Cohen, Kolatt, and Piran (1994) and at the nearly
3-$\sigma$ level reported by Rood and Struble (1996), although we found
that a 2.5-$\sigma$ excess was more representative.  We note that such a
correlation would not necessarily be expected from cosmological distances
to GRBs that have been proposed recently.  Rood and Struble (1996)
hypothesize that BATSE GRBs best correlated with ACO clusters with
redshifts of less than 0.1, while Wickramasinghe et al. (1993) and Fenimore
and Bloom (1995) find from fits of the GRB brightness distribution that
even the brightest GRBs in the BATSE sample would typically be at higher
redshifts than this.

In the above analysis, we found an increasingly strong correlation strength
when {\it all} ACO clusters were correlated with increasingly well located
GRBs. We found, however, no IPN or COMPTEL GRBs near these same ACO
clusters. This is slightly surprising because the IPN GRBs appear more
likely to correlate with ACO clusters than bright BATSE GRBs - the IPN GRBs
have higher peak flux and smaller locations.  The COMPTEL sample also has
higher peak flux, but had 7 GRBs in common with the BATSE sample, and in
some cases slightly larger error boxes. Even so, only a single random
coincident pair would have been expected, and perhaps as many as 3 real
ACO cluster/ IPN GRB pairs expected were the result reported by Rood and
Struble (1996) a true
mean value, given a linear extrapolation. So whereas the lack of any
coincident GRBs appears contrary to the previous claims, in reality it
carries little statistical weight.

We note that a simple calculation shows that the maximal ``3.5-$\sigma$"
excess in Figure 1 appears at least in part be due to statistical
fluctuation.  Even if every GRB of the best located 27 was associated with
an ACO cluster, the definition of the 1-$\sigma$ error box demands that, on
average, only 68.3 \% of them -- 18.4 in this case -- would be expected to
show an association, whereas 21 pairs were found.

It has been claimed, however, that the size of the published BATSE 3B
1-$\sigma$ error boxes were overestimated for the best located GRBs
(Graziani \& Lamb 1996). If true, each of the best located error boxes
would have a greater than 68.3 \% chance of housing an ACO cluster, which
again raises the possibility that the maximal ``3.5-$\sigma$" excess was
not due to statistical fluctuation.

Soft GRBs have a significantly different brightness distribution than the
unabridged set of BATSE GRBs (Belli 1991, Pizzichini 1994, Kouveliotou et al.
1996). In fact, the brightness distribution of the soft set is consistent
with a -1.5 power law, whereas the brightness distribution for the
unabridged set is NOT consistent with such a power law - it suffers from a
significant paucity of dim bursts.

The reality of marginal correlations like the ones presented here are
usually found untrue in light of better data.  The reason is usually {\it
not} that the computations were done incorrectly, but rather that only the
best results were presented.  We have made an effort here to present more
than just the best correlations. But there is always a dividing line where
the researchers are unsure whether the ``signal" they think they see is
real. In this light, and in the light of other recent claims of researchers
in this area, we chose to publish with the present disclaimers. An analogy
is this: while walking down the street a person thinks they hear a noise.
What might a person typically do to check whether the noise was real? They
might stop and listen more intently.  The marginal statistical
excesses we report above are reminiscent of a faint noise - we report them
here so the community can listen for them more intently in the future.

To test the reality of claims of correlations between clusters of galaxies
and GRBs, more data is needed, preferably of higher angular resolution. To
know exactly how many more GRBs are needed would require a model for how
GRBs and clusters of galaxies are associated. A ten-year BATSE lifetime may
well allow it to accumulate three times the number of GRBs analyzed in this
study. We eagerly look forward to the continuation of BATSE's mission!

\acknowledgments

This work was supported by a grant from NASA.

\clearpage

\figcaption[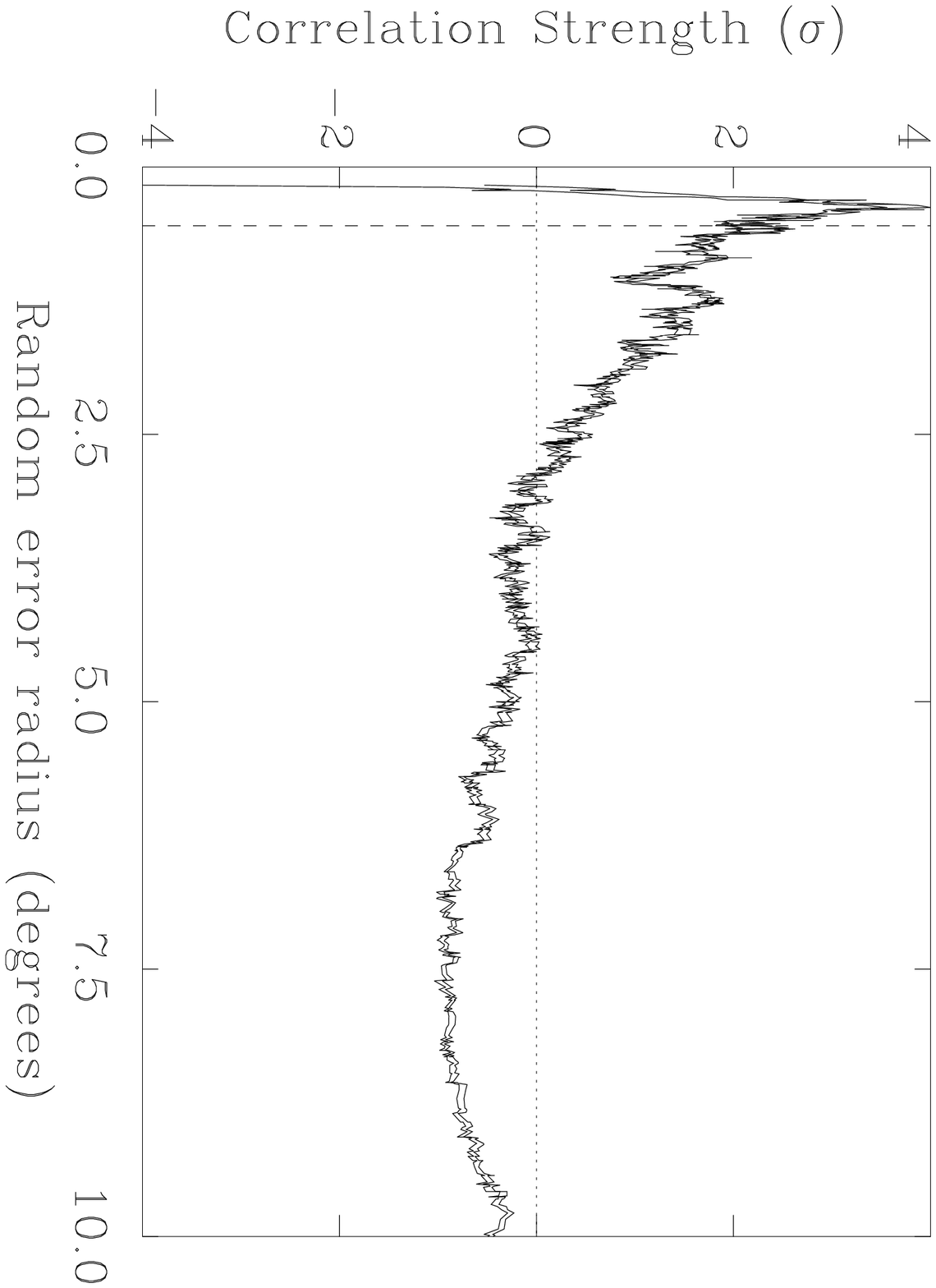]{A plot of correlation significance versus
GRB error box radius.  The correlation significance is computed between ACO
clusters and all BATSE 3B GRBs with random error box radii less than or
equal to the plotted value. We note correlation strength increases with
decreasing error box size.
\label{fig1}}

\figcaption[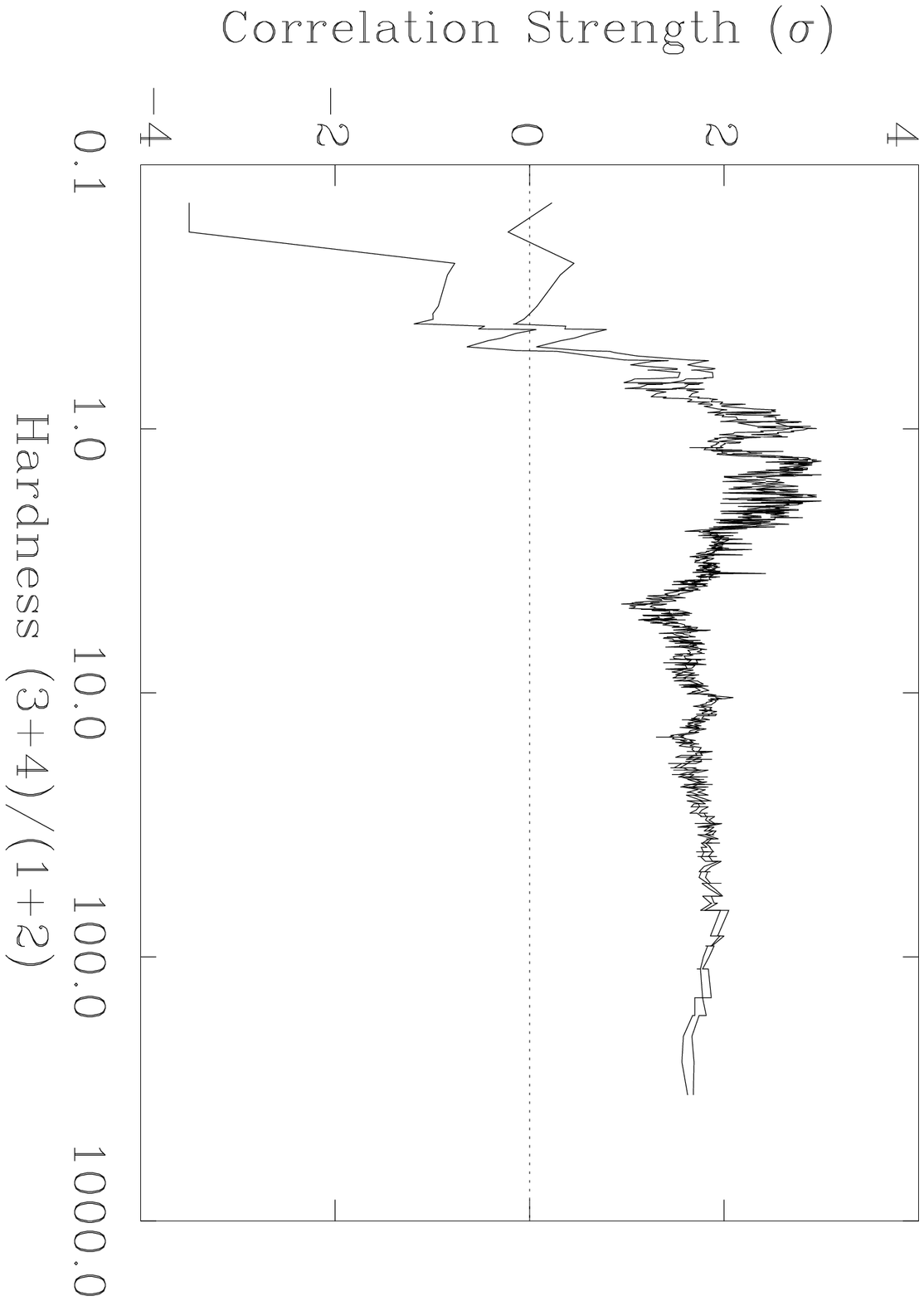]{A plot of correlation strength versus GRB
hardness ratio.  The correlation significance is computed between ACO
clusters and all BATSE 3B GRBs with hardness less than or equal to the
plotted value.  We note a slight correlation excess for soft GRBs.
\label{fig2}}

\end{document}